\documentclass[twocolumn,
prl
,superscriptaddress
]{revtex4}

\usepackage{graphicx}
\usepackage{amssymb}
\usepackage{amsmath}
\usepackage{color}
\usepackage{comment}

\newcommand{\ve}[1]{{\mathbf #1}}

\begin{document}


\title{Density-wave phases of dipolar fermions in a bilayer}

\author{F. M. Marchetti}
\affiliation{Departamento de F\'isica Te\'orica de la Materia
Condensada, Universidad Aut\'onoma de Madrid, Madrid 28049, Spain}

\author{M. M. Parish}
\affiliation{Cavendish Laboratory, JJ Thomson Avenue, Cambridge,
 CB3 0HE, United Kingdom} %
\affiliation{London Centre for Nanotechnology, Gordon Street, London, WC1H
0AH, United Kingdom}

\date{\today}

\begin{abstract}
  We investigate the phase diagram of dipolar fermions with aligned
  dipole moments in a two-dimensional (2D) bilayer.  Using a version
  of the Singwi-Tosi-Land-Sj\"{o}lander scheme recently adapted to
  dipolar fermions in a single layer [M. M. Parish and F.
  M. Marchetti, Phys. Rev. Lett. \textbf{108}, 145304 (2012)], we
  determine the density-wave instabilities of the bilayer system
  within linear response theory. We find that the bilayer geometry can
  stabilize the collapse of the 2D dipolar Fermi gas with intralayer
  attraction to form a new density wave phase that has an orientation
  perpendicular to the density wave expected for strong intralayer
  repulsion. We thus obtain a quantum phase transition between stripe
  phases that is driven by the interplay between strong correlations
  and the architecture of the low dimensional system.
\end{abstract}


\maketitle

Density-wave phases such as stripes are apparently ubiquitous in
nature. They are typically found in quasi-two-dimensional or layered
materials~\cite{CDW_review,Howald2003,graphene_stripe}, where they
manifest as periodic modulations of the electron density within the
two-dimensional (2D) layers. Moreover, such stripes have been linked
with high temperature
superconductivity~\cite{kivelson1998,kivelson2003}. However, despite
their ubiquity and potential importance, their origins and behavior
are still under debate. Indeed, a central question is whether stripes
are driven by electron-electron repulsion or simply by the
architecture of the underlying crystal structure~\cite{Mazin2008}.

One route to gaining insight into the problem is to study cleaner,
more tunable analogues of these electron systems. Quantum degenerate
Fermi gases with long-range dipolar
interactions~\cite{Baranov2008,Carr2009} provide just such a system in
which to investigate density-wave phases.  Such dipolar Fermi gases
have recently been realized experimentally with both magnetic
atoms~\cite{mingwu2012} and polar diatomic
molecules~\cite{Ni2008,heo2012,Wu2012}. In particular, ultracold polar
molecules of $^{40}$K $^{87}$Rb have been confined to 2D layers using
an optical lattice~\cite{miranda2011}, thus paving the way for
exploring long-range interactions in low dimensional systems.

For a 2D gas of polar molecules, the dipole-dipole interactions can be
controlled by aligning the dipole moments with an external electric
field. For small dipole tilt angles $\theta$ with respect to the plane
normal, the dipolar interactions are purely repulsive, while for
$\theta \gtrsim \pi/4$, the interactions acquire a significant
attractive component such that the dipolar Fermi system is unstable
towards collapse for sufficiently strong
interactions~\cite{bruun2008,yamaguchi2010,parish2012,sieberer2011}.
Away from collapse, in the repulsive regime, previous theoretical work
has predicted the existence of a stripe
phase~\cite{yamaguchi2010,parish2012,sieberer2011,babadi2011}, even
for the case where the dipolar interactions are \textit{isotropic}
($\theta=0$) and the system must spontaneously break rotational
symmetry~\cite{parish2012}. Here we investigate the effect of the low
dimensional architecture on density instabilities by considering
dipolar fermions in a 2D bilayer geometry.

\begin{figure}
\centering
\includegraphics[width=\linewidth,angle=0]{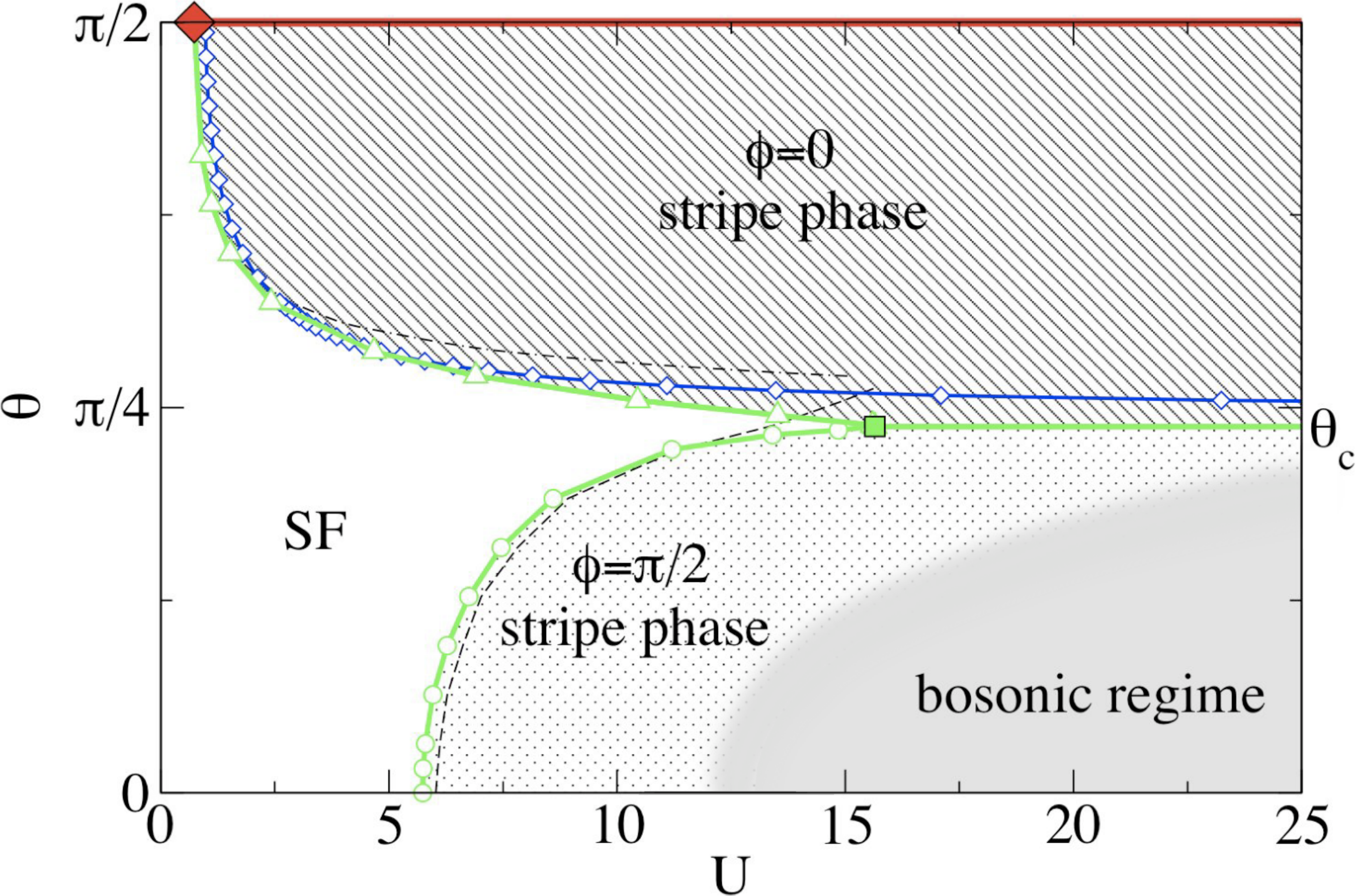}
\caption{(Color online) Phase diagram for a dipolar Fermi gas in a
  bilayer at fixed interlayer distance, $k_Fd = 2$, as a function of
  $\theta$ (see Fig.~\ref{fig:schematic}) and interaction
  $U=mD^2k_F/\hbar^2$.  The liquid phase is superfluid (SF).  The
  (green) open triangles [circles] set the boundary of the stripe
  phase oriented along $\phi = 0$ [$\phi=\pi/2$], derived from a
  self-consistent STLS calculation. The filled (green) square at
  $\theta_c \simeq 0.75$ and $U\simeq 15.65$ is a quantum critical
  point beyond which there is a phase transition between the two
  stripe phases. The (blue) open diamonds for the $\phi=0$ stripe
  phase are instead determined including exchange correlations only
  (see text). These boundaries can be compared to the $\phi=\pi/2$
  stripe transition (dashed line) and the collapse instability
  (dashed-dotted line) for the single-layer case~\cite{parish2012}.
  The shaded ``bosonic'' region is where the system can be described
  in terms of interlayer bosonic dimers. The (red) filled diamond and
  thick (red) line at $\theta=\pi/2$ indicate collapse in the
  bilayer.}
\label{fig:phase_diag}
\end{figure}

\begin{figure}
\centering
\includegraphics[width=\linewidth,angle=0]{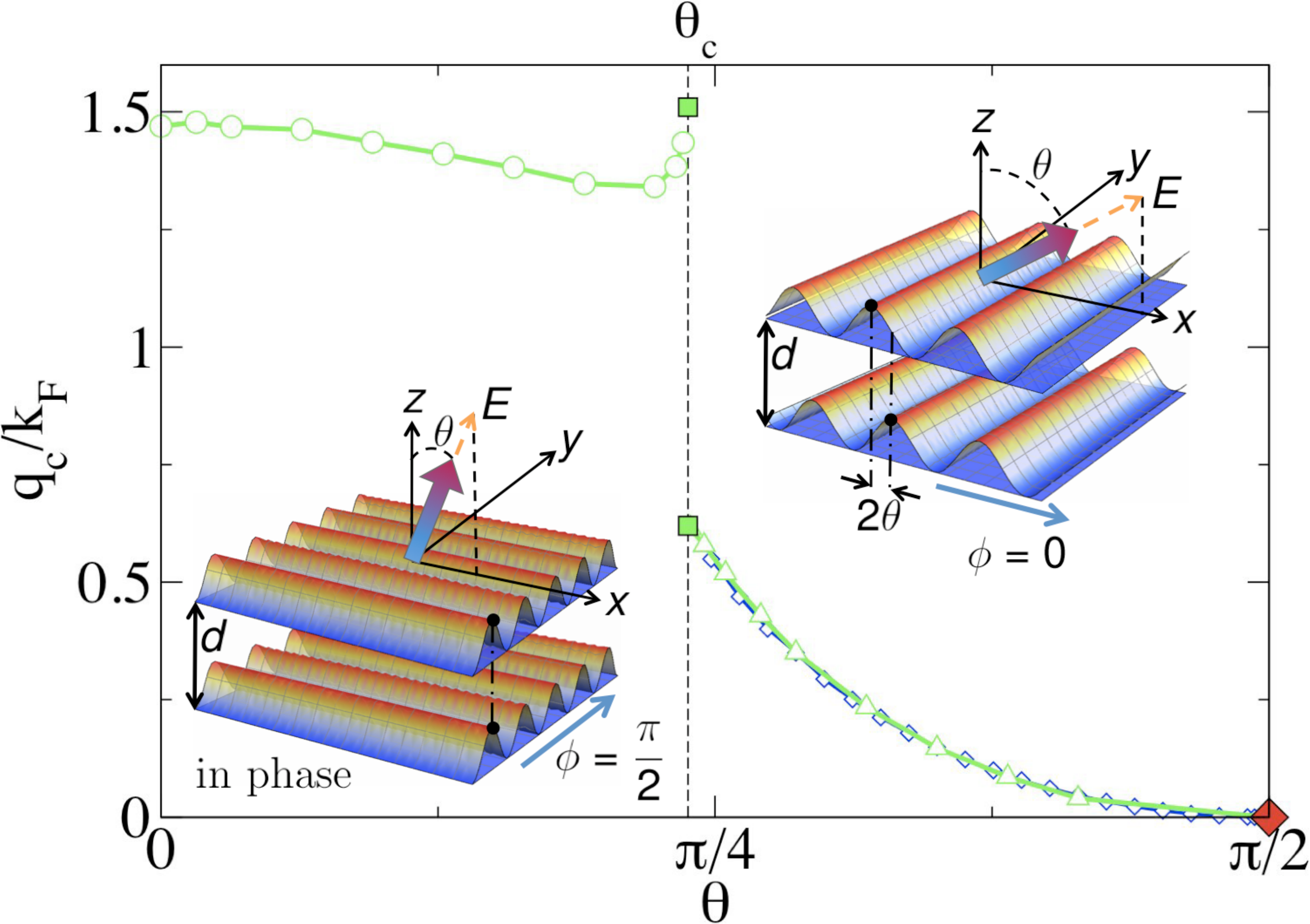}
\caption{(Color online) Critical wave vector $q_c/k_F$ for the
  $\phi=\pi/2$ stripe phase ($\theta < \theta_c$) and the $\phi = 0$
  one ($\theta > \theta_c$) --- same parameters and symbol scheme as
  in Fig.~\ref{fig:phase_diag}. The insets depict the alignment of the
  dipoles with the electric field $\ve{E}$ and the features of the two
  different stripe phases. For the $\phi = 0$ stripe phase, the
  density modulations in the two layers have a phase shift $\eta
  \simeq 2\theta$, while the wave vector $q_c$ decreases with
  increasing tilt angle $\theta$ down to $q_c=0$ for $\theta=\pi/2$
  (filled [red] diamond), where the gas collapses. For density
  modulations along $\phi = \pi/2$, $q_c$ appears to be fixed by the
  density.}
\label{fig:schematic}
\end{figure}

We determine the phase diagram of the bilayer system within linear
response theory, using a version of the Singwi-Tosi-Land-Sj\"{o}lander
(STLS) scheme~\cite{STLSpaper} recently developed in
Ref.~\cite{parish2012}. Based on this analysis, we show that the
bilayer geometry can actually stabilize the collapse of the 2D Fermi
gas to form a new density wave (Fig.~\ref{fig:phase_diag}). However,
in contrast to the stripes in the repulsive regime, this new stripe
phase has density modulations along the direction of the dipole tilt
(Fig.~\ref{fig:schematic}) and can also be well described by a
simplified STLS theory that involves exchange correlations only.  Our
work thus reveals a new quantum phase transition between two different
stripe modulations, where one phase is driven by strong repulsive
correlations and the other is driven by the bilayer architecture.

In the following, we consider the bilayer geometry shown in the insets
of Fig.~\ref{fig:schematic}.  Here, the dipole moments (of strength
$D$) are aligned by an external electric field $\ve{E}$ lying in the
$x$-$z$ plane and at angle $\theta$ with respect to the $z$
direction. We parameterize the $x$-$y$ in-plane momentum by polar
coordinates $\ve{q}=(q,\phi)$, with $\phi = 0$ corresponding to the
direction $x$ of the dipole tilt. The remaining system parameters are
the bilayer distance $d$, and the Fermi wave vector $k_F = \sqrt{4\pi
  n}$ ($n$ is the density in each layer). For dipoles confined in a
layer of width $W$, in the limit $q W\ll 1$, the effective 2D
intralayer interaction can be written as~\cite{fischer_06}:
\begin{equation}
  v_{11}(\ve{q}) = V_0 - 2\pi D^2 q \xi(\theta,\phi)\; ,
\label{eq:intra}
\end{equation}
where $\xi(\theta,\phi) = \cos^2\theta - \sin^2\theta \cos^2\phi$, and
$V_0$ is the $W$-dependent short-ranged contact interaction. The
confinement width $W$ provides a natural cut-off for the quasi-2D
system: $\Lambda \sim 1/W \gg k_F$.

Likewise, in the limit $W\ll d$, we can write the interlayer
interaction as~\cite{Li_Hwang_DasSarma10}:
\begin{equation}
  v_{12}(\ve{q}) =-2\pi D^2 q e^{-qd} \left[\xi(\theta,\phi) +i
    \sin2\theta \cos\phi \right]\; .
\label{eq:inter}
\end{equation}
Note that for $\theta \ne 0$, this interaction is complex and
satisfies $v_{21} (\ve{q}) = v_{12}^*(\ve{q}) = v_{12}(-\ve{q})$. This
arises from the fact that the interlayer interaction in real space is
not invariant under the transformation $\ve{r} \mapsto -\ve{r}$.

Assuming identical layers, one can parameterize the bilayer system
using only three dimensionless quantities: The tilt angle $\theta$,
the bilayer distance $k_F d$, and the interaction strength $U = mD^2
k_F/\hbar^2$, with $m$ being the fermion mass. The cut-off $\Lambda$
and the contact interaction $V_0$ should not be relevant since these
do not affect the low energy behavior of dipolar fermions, and indeed
the procedure we employ preserves this.

We now turn to the linear response theory used to analyze the
inhomogeneous phases of the dipolar system. In the bilayer (and
multilayers generally), the linear density response $\delta n_{i}$ to
an external perturbing field $V_{i}^{ext}$ defines the density-density
correlation function matrix $\chi_{ij}$,
\begin{equation}
  \delta n_{i} (\ve{q}, \omega)= \sum_j \chi_{ij} (\ve{q},\omega)
  V_{j}^{ext} (\ve{q},\omega)\; ,
\end{equation}
where $i$, $j$ are the layer indices. For a non-interacting gas, we
clearly have $\chi_{ij} = \delta_{ij} \Pi$, where the non-interacting
intralayer response function $\Pi(q,\omega)$ can be evaluated
analytically~\cite{stern67}.  Typically, one includes
interactions via the Random Phase Approximation (RPA), where one uses
a perturbing field that contains an effective potential due to the
perturbed density: $V_{j}^{ext} \mapsto V_{j}^{ext} + \sum_j
v_{ij}\delta n_{j}$, with intralayer potential $v_{22}(\ve{q}) =
v_{11}(\ve{q})$. However, as has been argued recently for the single
layer case, RPA is never accurate for dipolar interactions, since it
neglects exchange correlations~\cite{babadi2011,sieberer2011} which
are important even in the long-wavelength limit~\cite{parish2012}.

A straightforward and physically motivated way of incorporating
correlations beyond RPA is by means of local field factors
$G_{ij} (\ve{q})$ (for an introduction to this method see, e.g.,
Ref.~\cite{vignale_book}). Here, the (inverse) response function now
reads:
\begin{equation}
  {\chi^{-1}}_{ij} (\ve{q},\omega) = \frac{\delta_{ij}}{\Pi
    (q,\omega)} - v_{ij}(\ve{q}) \left[1 - G_{ij}(\ve{q}) \right]
  \; .  
\label{eq:response}
\end{equation}
Note that we clearly recover both RPA and the non-interacting case if
we take, respectively, $G_{ij}=0$ or $G_{ij}=1$. This response
function can be related to the ``layer-resolved'' static structure
factor $S_{ij}(\ve{q})$ by the fluctuation-dissipation theorem:
\begin{equation}
  S_{ij}(\ve{q}) = -\frac{\hbar}{\pi n} \int_0^{\infty} d\omega
  \chi_{ij}(\ve{q},i\omega)\; .
\label{eq:struc}
\end{equation}
In turn, we can approximate the local field factors using the STLS
scheme~\cite{STLSpaper}:
\begin{equation}
  G_{ij}(\ve{q}) = \frac{1}{n} \int \frac{d\ve{k}}{(2\pi)^2}
  \frac{\ve{q} \cdot \ve{k}}{q^2} \frac{v_{ij} (\ve{k})}{v_{ij}
    (\ve{q})} \left[ \delta_{ij} - S_{ij} (\ve{q}-\ve{k}) \right] \; .
\label{eq:local}
\end{equation}
The response function $\chi_{ij}$ (and associated structure factor
$S_{ij}$) can now be determined by solving
Eqs.~\eqref{eq:response}-\eqref{eq:local} self-consistently.  The STLS
scheme has been heavily utilized for Coulomb interactions and it has
proven to be very successful for describing the dielectric function of
several strongly-correlated electron systems (see~\cite{vignale_book}
and references therein). Following Ref.~\cite{parish2012}, we consider
an improved version of the STLS scheme that has been adapted to the
dipolar system. In essence, it ensures that our results are
insensitive to $\Lambda$ and $V_0$, by requiring that the intralayer
correlations be dominated by Pauli exclusion at large wavelengths $q
\gg 2k_F$.

For identical layers, we can assume that $S_{22}=S_{11}$,
$S_{21}=S_{12}^*$ (and similarly for the local field factors
$G_{ij}$). Note that the complex form of the interlayer
potential~\eqref{eq:inter} means that the interlayer factors $S_{12}
(\ve{q})$ and $G_{12}(\ve{q})$ are also complex. However, the symmetry
$v_{12}(-\ve{q})=v_{12}^*(\ve{q})$ is also preserved for both factors
at each iteration step of our self-consistent scheme. This guarantees
that physical quantities such as the ``layer-resolved'' pair
correlation functions, $g_{ij}(\ve{r}) = \frac{1}{n^2} \langle
\psi_i^\dag(\ve{r})\psi_j^\dag(0) \psi_j(0) \psi_i(\ve{r})\rangle$,
where
\begin{equation}
  g_{ij}(\ve{r}) = 1+ \frac{1}{n} \int \frac{d\ve{q}}{(2\pi)^2}
  e^{i\ve{q}.\ve{r}} \left[S_{ij}(\ve{q}) - \delta_{ij} \right]\; ,
\label{eq:pair}
\end{equation}
are always real, even when $i\neq j$. 

%

We determine the density instabilities of the bilayer system by
analyzing the divergences of the static response function matrix
$\chi_{ij} (\ve{q},0)$. Specifically, we search for zeros of the
largest inverse eigenvalue,
\begin{equation} 
  \chi_+^{-1} = \frac{1}{\Pi} - v_{11} [1-G_{11} ] + |v_{12} [1-G_{12}
  ] | \; . 
\label{eq:eigen} 
\end{equation}
A zero of $\chi_+^{-1} (\ve{q},0)$ at a critical wave vector
$\ve{q}_c$ signals an instability towards the formation of a density
wave with period set by $\ve{q_c}$.  If the instability occurs for a
specific direction $\phi$, then the density-wave phase corresponds to
a one-dimensional modulation (or stripe phase) of period $2\pi/q_c$
oriented along $\phi$. In this way, we obtain the phase diagram
plotted in Fig.~\ref{fig:phase_diag} for $k_F d=2$.

For tilt angles $\theta<\theta_c \simeq 0.75$, we find a stripe phase
along $\phi=\pi/2$ that is of a similar nature to the one found in a
single layer (dashed line of Fig.~\ref{fig:phase_diag}). In
particular, it is driven by strong intralayer correlations induced by
the repulsive part of $v_{11}$, as evidenced by the relative
insensitivity of $q_c$ to the bilayer geometry and $\theta$ (see
Fig.~\ref{fig:schematic}). However, the presence of the second layer
can decrease the value of the critical interaction strength $U_c$ for
stripe formation, as one might expect from the form of
Eq.~\eqref{eq:eigen}. The attractive part of $v_{12}(\ve{q})$ also
ensures that the density waves along $\phi=\pi/2$ in each layer are in
phase.
Similar results were found using the conserving Hartree-Fock (HF)
approximation~\cite{babadi2011,block2012}, but for much smaller values
of $U_c$, like in the single-layer case.
The shift of $U_c$ due to the other layer is relatively small for
distance $k_F d=2$ (see Fig.~\ref{fig:phase_diag} at small values of
$\theta$), but it can become substantial for smaller $k_F d$ since
Eq.~\eqref{eq:eigen} depends exponentially on the bilayer distance.
However, for smaller distances, we then encounter phases involving
strong interlayer pairing~\cite{pikovski2010,zinner_10,baranov2011}
and the system would instead be better described in terms of
interlayer bosonic dimers, as we discuss later.

In the isotropic case ($\theta=0$), we find that the system
spontaneously breaks rotational symmetry to form a stripe phase at $U
\simeq 5.74$, similarly to the single-layer case~\cite{parish2012}.
One can only observe this symmetry breaking at $\theta=0$ by starting
the STLS iteration with a solution for small but finite $\theta$. This
effectively corresponds to taking the limit $\theta \to 0$, which is
somewhat akin to classical ferromagnetism, where one must consider the
limit where magnetic field goes to zero.
This stripe phase precedes Wigner crystallization which, according to
quantum Monte Carlo (QMC) calculations, occurs at $U\simeq 25$ for
perpendicular fermionic dipoles in a single layer~\cite{matveeva2012}.

For $\theta>\arcsin(1/\sqrt{3})$, the intralayer interaction develops
an attractive sliver in the plane that can eventually lead to collapse
in the single
layer~\cite{bruun2008,yamaguchi2010,sieberer2011,parish2012}.  Here,
for large enough $U$ and $\theta$, the attraction overcomes Pauli
exclusion and the compressibility of the gas goes to zero
($\chi_+^{-1} (\ve{q}\to 0,0)=0$). However, we find that the bilayer
geometry can actually stabilize the collapse to form a new
density-wave phase that is oriented along the $\phi=0$ direction
(Fig.~\ref{fig:phase_diag}). Referring to Fig.~\ref{fig:schematic}, we
see that this stripe phase has a longer wavelength than the
$\phi=\pi/2$ one and is dependent on geometry. Indeed, we find that
$q_c$ smoothly decreases with increasing $\theta$, reaching $q_c=0$ at
$\theta=\pi/2$, where the intralayer attraction always appears to
cause collapse at a fixed $U_c$.
Away from $\theta = \pi/2$, we find that the $\phi=0$ stripe phase has
$q_c \sim 1/d$ in the limit $d \to \infty$, which is reminiscent of
the behavior of charge density waves in electron-hole bilayers. 

The $\phi=0$ stripe also features a nontrivial phase shift $\eta$
between the density waves in each layer. At the stripe transition, it
can be shown that
\begin{equation}
  e^{i\eta} = - \frac{v_{12} (\ve{q}) [1-G_{12}(\ve{q})]}{|v_{12}
    (\ve{q}) [1-G_{12}(\ve{q})]|}\; .
\end{equation}
When $v_{12}$ and $G_{12}$ are real, like for the $\phi=\pi/2$ stripe
phase, then $e^{i\eta} = 1$ and the density waves in each layer are in
phase, as mentioned previously. However, $v_{12}$ is complex for the
$\phi=0$ stripe phase and thus the density waves are generally shifted
with respect to one another. Indeed, as shown below, the interlayer
correlations are small in this phase, i.e.\ $|G_{12}|\ll 1$, therefore
the phase shift corresponds to $\eta \simeq 2\theta$ (see insets of
Fig.~\ref{fig:schematic}) and is essentially independent of $k_F d$.

The existence of two stripe phases leads to a new quantum phase
transition where the stripes change their orientation. In
Fig.~\ref{fig:phase_diag}, this occurs beyond the critical point
$\theta_c \simeq 0.75$ and $U_c \simeq 15.65$ where the two stripe
phase boundaries meet.
Here, when $k_F d$ is fixed, the transition can be accessed by
changing the tilt angle $\theta$. Alternatively, one can fix $\theta
\lesssim \pi/4$, which is below the onset of collapse in the single
layer, and vary $k_F d$, since we expect the critical angle $\theta_c$
to decrease with decreasing $k_F d$.  Eventually, at $k_F d \simeq 1$,
one enters the regime where the physics of bosonic interlayer dimers
dominates.

\begin{figure}
\centering
\includegraphics[width=0.9\linewidth,angle=0]{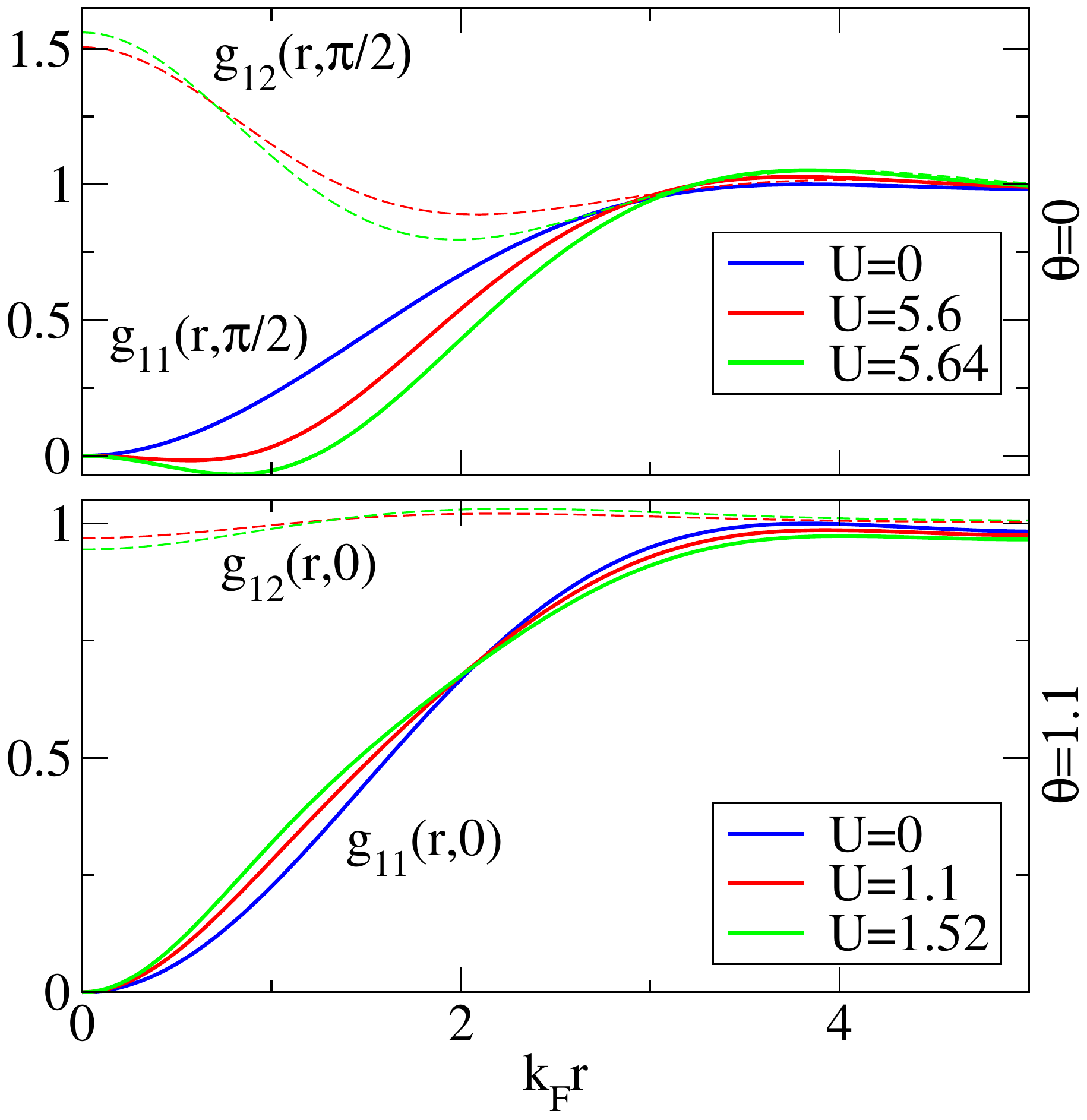}
\caption{(Color online) Intra- and interlayer pair correlation
  functions $g_{ij} (\ve{r})$ for increasing values of the interaction
  strength $U$ towards the $\phi = \pi/2$ stripe phase ($\theta=0$ top
  panel) and the $\phi = 0$ phase ($\theta=1.1 \simeq 0.35\pi$ bottom
  panel).}
\label{fig:correlation}
\end{figure}

Further insight into the stripe phases can be gained by examining the
intra- and interlayer pair correlation functions $g_{ij}(\ve{r})$ on
the liquid side of the transition. For the $\phi=0$ stripe phase
(bottom panel of Fig.~\ref{fig:correlation}), we find that neither
pair correlation function changes significantly as we approach the
transition.  In particular, $g_{11}(\ve{r})$ only deviates slightly
from the non-interacting case ($U=0$), while $g_{12}(\ve{r})$ slowly
oscillates close to one, indicating that interlayer correlations are
small, i.e.\ $|G_{12}|\ll 1$.
This suggests that we can accurately model the $\phi=0$ stripe phase
using exchange correlations only. To this end, we construct
a simplified STLS theory where we take $G_{12}(\ve{q}) = 0$ and then
determine the intralayer local field factor $G_{11} (\ve{q})$ by
feeding the non-interacting intralayer structure factor $S_{0}(q)
  = -\frac{\hbar}{\pi n} \int_0^{\infty} d\omega \Pi(q,\omega)$
%
%
into Eq.~\eqref{eq:local}. We then evaluate
the phase boundary for the $\phi=0$ stripe within this simplified HF
theory. Referring to Figs.~\ref{fig:phase_diag} and
\ref{fig:schematic}, we see that we obtain very good agreement with
the full STLS calculation, particularly when $U$ and $\theta$ are not
too large so that the intralayer $p$-wave pairing correlations are
expected to be weakest~\cite{bruun2008,sieberer2011}.
In addition, the collapse instability at $\theta = \pi/2$ is
unaffected by the other layer since the interlayer Hartree term is
zero for $\ve{q}=0$.
We expect one can obtain quantitatively similar results for the
$\phi=0$ stripe phase using the conserving HF
approximation~\footnote{Note that the $\phi=0$ stripe phase was not
  observed in Ref.~\cite{block2012} since they focused on the
  $\phi=\pi/2$ instability and ignored the imaginary part of
  $v_{12}(\ve{q})$.}.

By contrast, for the $\phi = \pi/2$ stripe phase (top panel of
Fig.~\ref{fig:correlation}), we see that correlations beyond exchange
become substantial, resulting in a pronounced ``correlation hole'' for
$g_{11} (\ve{r})$ with increasing interaction strength, like in the
single-layer case~\cite{parish2012} --- note that the STLS procedure
does not guarantee that $g_{11}$ is always
positive~\cite{vignale_book}, and thus we sometimes obtain unphysical
negative values.
The intralayer correlations also develop a substantial $\phi$
anisotropy as we near the stripe transition. At the same time, the
interlayer pair correlation function $g_{12}(\ve{r})$ increases at
$\ve{r}=0$, a feature that has been ascribed to an imminent
bound-state instability~\cite{Liu1998}.

Indeed, the attractive part of $v_{12}(\ve{q})$ always yields a
two-body bound state composed of one fermion from each
layer~\cite{klawunn2010,volosniev2011}. Hence, any liquid phase in the
phase diagram contains pairing correlations and must therefore be
superfluid (Fig.~\ref{fig:phase_diag}).
When the size of these interlayer dimers $l_B$ is smaller than the
interparticle spacing, i.e.\ $l_B\ll 1/k_F$, then the system is better
described in terms of bosonic dimers and our approach of analyzing
density instabilities of the Fermi liquid phase is unlikely to be
accurate. To estimate this region of phase space where bosonic
behavior dominates, we solve the two-body problem, $E \psi_{\ve{k}} =
\frac{\hbar^2 \ve{k}^2}{m} \psi_{\ve{k}} + \int
\frac{d\ve{k}}{(2\pi)^2} v_{12} (\ve{k}-\ve{k'}) \psi_{\ve{k'}}$,
where $\psi_{\ve{k}}$ is the two-body wave function in terms of
relative coordinates and $E$ is the dimer binding energy. We estimate
the dimer size as $l_B \sim \hbar/\sqrt{m|E|}$ and then determine the
``critical'' line $k_F l_B = 1$ for the bosonic regime, as plotted in
Fig.~\ref{fig:phase_diag} (shaded region). We see that this region is
well separated from the stripe phase boundaries and thus we expect our
results to be reasonable for $k_F d =2$.  However, the presence of
bosonic dimers hastens the onset Wigner crystallization: QMC
calculations~\cite{astrakharchik_07,buchler2007} predict that
perpendicularly-aligned bosons will crystallize at $U\simeq 8$.
For increasing $\theta$, the interlayer dimer becomes more weakly
bound until eventually the fermions preferentially form pairs within
the same layer instead.
With decreasing $k_F d$, however, the regime of interlayer bosons
expands so that it encroaches on our predicted stripe transitions for
$k_Fd \simeq 1$ and takes us beyond the scope of this letter.
%

Our predicted stripe phases should be accessible experimentally 
with cold dipolar gases. In particular, the bilayer distance $k_F d=2$ can be achieved
  for a typical 2D density $n \sim 1.3 \times 10^8$~cm$^{-2}$ and layer spacing 
 $d=500$~nm. Polar molecules such as LiCs~\cite{Carr2009} have dipolar moments  
 $D\sim 0.35-1.3$~Debye (corresponding to $U\sim 1-14$),
  which allows one to explore both $\phi=0$ and $\phi=\pi/2$ stripe
  phases. Furthermore, the newly explored NaK
  molecules~\cite{Wu2012} allows one to reach even larger values
  of the interaction strength ($D\sim 2.7$~Debye and $U\sim 28$).

\acknowledgments We are grateful to J. Levinsen,
P. Littlewood, and N. Zinner for useful discussions. MMP acknowledges
support from the EPSRC under Grant No.\ EP/H00369X/1. FMM acknowledges
financial support from the programs Ram\'on y Cajal and Intelbiomat
(ESF). We also acknowledge TCM group (Cambridge) for hospitality.
%

\bibliography{dipoleRefs}

\begin{thebibliography}{33}
\expandafter\ifx\csname natexlab\endcsname\relax\def\natexlab#1{#1}\fi
\expandafter\ifx\csname bibnamefont\endcsname\relax
  \def\bibnamefont#1{#1}\fi
\expandafter\ifx\csname bibfnamefont\endcsname\relax
  \def\bibfnamefont#1{#1}\fi
\expandafter\ifx\csname citenamefont\endcsname\relax
  \def\citenamefont#1{#1}\fi
\expandafter\ifx\csname url\endcsname\relax
  \def\url#1{\texttt{#1}}\fi
\expandafter\ifx\csname urlprefix\endcsname\relax\def\urlprefix{URL }\fi
\providecommand{\bibinfo}[2]{#2}
\providecommand{\eprint}[2][]{\url{#2}}

\bibitem[{\citenamefont{Gr\"uner}(1988)}]{CDW_review}
\bibinfo{author}{\bibfnamefont{G.}~\bibnamefont{Gr\"uner}},
  \bibinfo{journal}{Rev. Mod. Phys.} \textbf{\bibinfo{volume}{60}},
  \bibinfo{pages}{1129} (\bibinfo{year}{1988}).

\bibitem[{\citenamefont{Howald et~al.}(2003)\citenamefont{Howald, Eisaki,
  Kaneko, and Kapitulnik}}]{Howald2003}
\bibinfo{author}{\bibfnamefont{C.}~\bibnamefont{Howald}},
  \bibinfo{author}{\bibfnamefont{H.}~\bibnamefont{Eisaki}},
  \bibinfo{author}{\bibfnamefont{N.}~\bibnamefont{Kaneko}}, \bibnamefont{and}
  \bibinfo{author}{\bibfnamefont{A.}~\bibnamefont{Kapitulnik}},
  \bibinfo{journal}{Proc. Natl Acad. Sci. USA} \textbf{\bibinfo{volume}{100}},
  \bibinfo{pages}{9705} (\bibinfo{year}{2003}).

\bibitem[{\citenamefont{Rahnejat et~al.}(2011)\citenamefont{Rahnejat, Howard,
  Shuttleworth, S.~R. Schofield~and, Hirjibehedin, Renner, Aeppli, and
  Ellerby}}]{graphene_stripe}
\bibinfo{author}{\bibfnamefont{K.~C.} \bibnamefont{Rahnejat}},
  \bibinfo{author}{\bibfnamefont{C.~A.} \bibnamefont{Howard}},
  \bibinfo{author}{\bibfnamefont{N.~E.} \bibnamefont{Shuttleworth}},
  \bibinfo{author}{\bibfnamefont{K.~I.} \bibnamefont{S.~R. Schofield~and}},
  \bibinfo{author}{\bibfnamefont{C.~F.} \bibnamefont{Hirjibehedin}},
  \bibinfo{author}{\bibfnamefont{C.}~\bibnamefont{Renner}},
  \bibinfo{author}{\bibfnamefont{G.}~\bibnamefont{Aeppli}}, \bibnamefont{and}
  \bibinfo{author}{\bibfnamefont{M.}~\bibnamefont{Ellerby}},
  \bibinfo{journal}{Nat. Commun.} \textbf{\bibinfo{volume}{2}},
  \bibinfo{pages}{558} (\bibinfo{year}{2011}).

\bibitem[{\citenamefont{Kivelson et~al.}(1998)\citenamefont{Kivelson, Fradkin,
  and Emery}}]{kivelson1998}
\bibinfo{author}{\bibfnamefont{S.~A.} \bibnamefont{Kivelson}},
  \bibinfo{author}{\bibfnamefont{E.}~\bibnamefont{Fradkin}}, \bibnamefont{and}
  \bibinfo{author}{\bibfnamefont{V.~J.} \bibnamefont{Emery}},
  \bibinfo{journal}{Nature} \textbf{\bibinfo{volume}{393}},
  \bibinfo{pages}{550} (\bibinfo{year}{1998}).

\bibitem[{\citenamefont{Kivelson et~al.}(2003)\citenamefont{Kivelson, Bindloss,
  Fradkin, Oganesyan, Tranquada, Kapitulnik, and Howald}}]{kivelson2003}
\bibinfo{author}{\bibfnamefont{S.~A.} \bibnamefont{Kivelson}},
  \bibinfo{author}{\bibfnamefont{I.~P.} \bibnamefont{Bindloss}},
  \bibinfo{author}{\bibfnamefont{E.}~\bibnamefont{Fradkin}},
  \bibinfo{author}{\bibfnamefont{V.}~\bibnamefont{Oganesyan}},
  \bibinfo{author}{\bibfnamefont{J.~M.} \bibnamefont{Tranquada}},
  \bibinfo{author}{\bibfnamefont{A.}~\bibnamefont{Kapitulnik}},
  \bibnamefont{and} \bibinfo{author}{\bibfnamefont{C.}~\bibnamefont{Howald}},
  \bibinfo{journal}{Rev. Mod. Phys.} \textbf{\bibinfo{volume}{75}},
  \bibinfo{pages}{1201} (\bibinfo{year}{2003}).

\bibitem[{\citenamefont{Johannes and Mazin}(2008)}]{Mazin2008}
\bibinfo{author}{\bibfnamefont{M.~D.} \bibnamefont{Johannes}} \bibnamefont{and}
  \bibinfo{author}{\bibfnamefont{I.~I.} \bibnamefont{Mazin}},
  \bibinfo{journal}{Phys. Rev. B} \textbf{\bibinfo{volume}{77}},
  \bibinfo{pages}{165135} (\bibinfo{year}{2008}).

\bibitem[{\citenamefont{Baranov}(2008)}]{Baranov2008}
\bibinfo{author}{\bibfnamefont{M.~A.} \bibnamefont{Baranov}},
  \bibinfo{journal}{Phys. Rep.} \textbf{\bibinfo{volume}{464}},
  \bibinfo{pages}{71 } (\bibinfo{year}{2008}).

\bibitem[{\citenamefont{Carr et~al.}(2009)\citenamefont{Carr, DeMille, Krems,
  and Ye}}]{Carr2009}
\bibinfo{author}{\bibfnamefont{L.~D.} \bibnamefont{Carr}},
  \bibinfo{author}{\bibfnamefont{D.}~\bibnamefont{DeMille}},
  \bibinfo{author}{\bibfnamefont{R.~V.} \bibnamefont{Krems}}, \bibnamefont{and}
  \bibinfo{author}{\bibfnamefont{J.}~\bibnamefont{Ye}}, \bibinfo{journal}{New
  J. Phys.} \textbf{\bibinfo{volume}{11}}, \bibinfo{pages}{055049}
  (\bibinfo{year}{2009}).

\bibitem[{\citenamefont{Lu et~al.}()\citenamefont{Lu, Burdick, and
  Lev}}]{mingwu2012}
\bibinfo{author}{\bibfnamefont{M.}~\bibnamefont{Lu}},
  \bibinfo{author}{\bibfnamefont{N.~Q.} \bibnamefont{Burdick}},
  \bibnamefont{and} \bibinfo{author}{\bibfnamefont{B.~L.} \bibnamefont{Lev}},
  \bibinfo{note}{arXiv:1202.4444}.

\bibitem[{\citenamefont{Ni et~al.}(2008)\citenamefont{Ni, Ospelkaus,
  De~Miranda, Pe'er, Neyenhuis, Zirbel, Kotochigova, Julienne, Jin, and
  Ye}}]{Ni2008}
\bibinfo{author}{\bibfnamefont{K.~K.} \bibnamefont{Ni}},
  \bibinfo{author}{\bibfnamefont{S.}~\bibnamefont{Ospelkaus}},
  \bibinfo{author}{\bibfnamefont{M.~H.~G.} \bibnamefont{De~Miranda}},
  \bibinfo{author}{\bibfnamefont{A.}~\bibnamefont{Pe'er}},
  \bibinfo{author}{\bibfnamefont{B.}~\bibnamefont{Neyenhuis}},
  \bibinfo{author}{\bibfnamefont{J.~J.} \bibnamefont{Zirbel}},
  \bibinfo{author}{\bibfnamefont{S.}~\bibnamefont{Kotochigova}},
  \bibinfo{author}{\bibfnamefont{P.~S.} \bibnamefont{Julienne}},
  \bibinfo{author}{\bibfnamefont{D.~S.} \bibnamefont{Jin}}, \bibnamefont{and}
  \bibinfo{author}{\bibfnamefont{J.}~\bibnamefont{Ye}},
  \bibinfo{journal}{Science} \textbf{\bibinfo{volume}{322}},
  \bibinfo{pages}{231} (\bibinfo{year}{2008}).

\bibitem[{\citenamefont{Heo et~al.}()\citenamefont{Heo, Wang, Christensen,
  Rvachov, Cotta, Choi, Lee, and Ketterle}}]{heo2012}
\bibinfo{author}{\bibfnamefont{M.-S.} \bibnamefont{Heo}},
  \bibinfo{author}{\bibfnamefont{T.~T.} \bibnamefont{Wang}},
  \bibinfo{author}{\bibfnamefont{C.~A.} \bibnamefont{Christensen}},
  \bibinfo{author}{\bibfnamefont{T.~M.} \bibnamefont{Rvachov}},
  \bibinfo{author}{\bibfnamefont{D.~A.} \bibnamefont{Cotta}},
  \bibinfo{author}{\bibfnamefont{J.-H.} \bibnamefont{Choi}},
  \bibinfo{author}{\bibfnamefont{Y.-R.} \bibnamefont{Lee}}, \bibnamefont{and}
  \bibinfo{author}{\bibfnamefont{W.}~\bibnamefont{Ketterle}},
  \bibinfo{note}{arXiv:1205.5304}.

\bibitem[{\citenamefont{Wu et~al.}()\citenamefont{Wu, Park, Ahmadi, Will, and
  Zwierlein}}]{Wu2012}
\bibinfo{author}{\bibfnamefont{C.-H.} \bibnamefont{Wu}},
  \bibinfo{author}{\bibfnamefont{J.~W.} \bibnamefont{Park}},
  \bibinfo{author}{\bibfnamefont{P.}~\bibnamefont{Ahmadi}},
  \bibinfo{author}{\bibfnamefont{S.}~\bibnamefont{Will}}, \bibnamefont{and}
  \bibinfo{author}{\bibfnamefont{M.~W.} \bibnamefont{Zwierlein}},
  \bibinfo{note}{arXiv:1206.5023}.

\bibitem[{\citenamefont{de~Miranda et~al.}(2011)\citenamefont{de~Miranda,
  Chotia, Neyenhuis, Wang, Qu\'{e}m\'{e}ner, Ospelkaus, Bohn, Ye, and
  Jin}}]{miranda2011}
\bibinfo{author}{\bibfnamefont{M.~H.~G.} \bibnamefont{de~Miranda}},
  \bibinfo{author}{\bibfnamefont{A.}~\bibnamefont{Chotia}},
  \bibinfo{author}{\bibfnamefont{B.}~\bibnamefont{Neyenhuis}},
  \bibinfo{author}{\bibfnamefont{D.}~\bibnamefont{Wang}},
  \bibinfo{author}{\bibfnamefont{G.}~\bibnamefont{Qu\'{e}m\'{e}ner}},
  \bibinfo{author}{\bibfnamefont{S.}~\bibnamefont{Ospelkaus}},
  \bibinfo{author}{\bibfnamefont{J.~L.} \bibnamefont{Bohn}},
  \bibinfo{author}{\bibfnamefont{J.}~\bibnamefont{Ye}}, \bibnamefont{and}
  \bibinfo{author}{\bibfnamefont{D.~S.} \bibnamefont{Jin}},
  \bibinfo{journal}{Nature Phys.} \textbf{\bibinfo{volume}{7}},
  \bibinfo{pages}{502} (\bibinfo{year}{2011}).

\bibitem[{\citenamefont{Bruun and Taylor}(2008)}]{bruun2008}
\bibinfo{author}{\bibfnamefont{G.~M.} \bibnamefont{Bruun}} \bibnamefont{and}
  \bibinfo{author}{\bibfnamefont{E.}~\bibnamefont{Taylor}},
  \bibinfo{journal}{Phys. Rev. Lett.} \textbf{\bibinfo{volume}{101}},
  \bibinfo{pages}{245301} (\bibinfo{year}{2008}).

\bibitem[{\citenamefont{Yamaguchi et~al.}(2010)\citenamefont{Yamaguchi, Sogo,
  Ito, and Miyakawa}}]{yamaguchi2010}
\bibinfo{author}{\bibfnamefont{Y.}~\bibnamefont{Yamaguchi}},
  \bibinfo{author}{\bibfnamefont{T.}~\bibnamefont{Sogo}},
  \bibinfo{author}{\bibfnamefont{T.}~\bibnamefont{Ito}}, \bibnamefont{and}
  \bibinfo{author}{\bibfnamefont{T.}~\bibnamefont{Miyakawa}},
  \bibinfo{journal}{Phys. Rev. A} \textbf{\bibinfo{volume}{82}},
  \bibinfo{pages}{013643} (\bibinfo{year}{2010}).

\bibitem[{\citenamefont{Parish and Marchetti}(2012)}]{parish2012}
\bibinfo{author}{\bibfnamefont{M.~M.} \bibnamefont{Parish}} \bibnamefont{and}
  \bibinfo{author}{\bibfnamefont{F.~M.} \bibnamefont{Marchetti}},
  \bibinfo{journal}{Phys. Rev. Lett.} \textbf{\bibinfo{volume}{108}},
  \bibinfo{pages}{145304} (\bibinfo{year}{2012}).

\bibitem[{\citenamefont{Sieberer and Baranov}(2011)}]{sieberer2011}
\bibinfo{author}{\bibfnamefont{L.~M.} \bibnamefont{Sieberer}} \bibnamefont{and}
  \bibinfo{author}{\bibfnamefont{M.~A.} \bibnamefont{Baranov}},
  \bibinfo{journal}{Phys. Rev. A} \textbf{\bibinfo{volume}{84}},
  \bibinfo{pages}{063633} (\bibinfo{year}{2011}).

\bibitem[{\citenamefont{Babadi and Demler}(2011)}]{babadi2011}
\bibinfo{author}{\bibfnamefont{M.}~\bibnamefont{Babadi}} \bibnamefont{and}
  \bibinfo{author}{\bibfnamefont{E.}~\bibnamefont{Demler}},
  \bibinfo{journal}{Phys. Rev. B} \textbf{\bibinfo{volume}{84}},
  \bibinfo{pages}{235124} (\bibinfo{year}{2011}).

\bibitem[{\citenamefont{Singwi et~al.}(1968)\citenamefont{Singwi, Tosi, Land,
  and Sj\"olander}}]{STLSpaper}
\bibinfo{author}{\bibfnamefont{K.~S.} \bibnamefont{Singwi}},
  \bibinfo{author}{\bibfnamefont{M.~P.} \bibnamefont{Tosi}},
  \bibinfo{author}{\bibfnamefont{R.~H.} \bibnamefont{Land}}, \bibnamefont{and}
  \bibinfo{author}{\bibfnamefont{A.}~\bibnamefont{Sj\"olander}},
  \bibinfo{journal}{Phys. Rev.} \textbf{\bibinfo{volume}{176}},
  \bibinfo{pages}{589} (\bibinfo{year}{1968}).

\bibitem[{\citenamefont{Fischer}(2006)}]{fischer_06}
\bibinfo{author}{\bibfnamefont{U.~R.} \bibnamefont{Fischer}},
  \bibinfo{journal}{Phys. Rev. A} \textbf{\bibinfo{volume}{73}},
  \bibinfo{pages}{031602} (\bibinfo{year}{2006}).

\bibitem[{\citenamefont{Li et~al.}(2010)\citenamefont{Li, Hwang, and
  Das~Sarma}}]{Li_Hwang_DasSarma10}
\bibinfo{author}{\bibfnamefont{Q.}~\bibnamefont{Li}},
  \bibinfo{author}{\bibfnamefont{E.~H.} \bibnamefont{Hwang}}, \bibnamefont{and}
  \bibinfo{author}{\bibfnamefont{S.}~\bibnamefont{Das~Sarma}},
  \bibinfo{journal}{Phys. Rev. B} \textbf{\bibinfo{volume}{82}},
  \bibinfo{pages}{235126} (\bibinfo{year}{2010}).

\bibitem[{\citenamefont{Stern}(1967)}]{stern67}
\bibinfo{author}{\bibfnamefont{F.}~\bibnamefont{Stern}},
  \bibinfo{journal}{Phys. Rev. Lett.} \textbf{\bibinfo{volume}{18}},
  \bibinfo{pages}{546} (\bibinfo{year}{1967}).

\bibitem[{\citenamefont{Giuliani and Vignale}(2005)}]{vignale_book}
\bibinfo{author}{\bibfnamefont{G.~F.} \bibnamefont{Giuliani}} \bibnamefont{and}
  \bibinfo{author}{\bibfnamefont{G.}~\bibnamefont{Vignale}},
  \emph{\bibinfo{title}{Quantum Theory of the Electron Liquid}}
  (\bibinfo{publisher}{Cambridge University Press}, \bibinfo{year}{2005}).

\bibitem[{\citenamefont{Block et~al.}()\citenamefont{Block, Zinner, and
  Bruun}}]{block2012}
\bibinfo{author}{\bibfnamefont{J.~K.} \bibnamefont{Block}},
  \bibinfo{author}{\bibfnamefont{N.~T.} \bibnamefont{Zinner}},
  \bibnamefont{and} \bibinfo{author}{\bibfnamefont{G.~M.} \bibnamefont{Bruun}},
  \bibinfo{note}{arXiv:1204.1822}.

\bibitem[{\citenamefont{Pikovski et~al.}(2010)\citenamefont{Pikovski, Klawunn,
  Shlyapnikov, and Santos}}]{pikovski2010}
\bibinfo{author}{\bibfnamefont{A.}~\bibnamefont{Pikovski}},
  \bibinfo{author}{\bibfnamefont{M.}~\bibnamefont{Klawunn}},
  \bibinfo{author}{\bibfnamefont{G.~V.} \bibnamefont{Shlyapnikov}},
  \bibnamefont{and} \bibinfo{author}{\bibfnamefont{L.}~\bibnamefont{Santos}},
  \bibinfo{journal}{Phys. Rev. Lett.} \textbf{\bibinfo{volume}{105}},
  \bibinfo{pages}{215302} (\bibinfo{year}{2010}).

\bibitem[{\citenamefont{Zinner et~al.}(2012)\citenamefont{Zinner, Wunsch,
  Pekker, and Wang}}]{zinner_10}
\bibinfo{author}{\bibfnamefont{N.~T.} \bibnamefont{Zinner}},
  \bibinfo{author}{\bibfnamefont{B.}~\bibnamefont{Wunsch}},
  \bibinfo{author}{\bibfnamefont{D.}~\bibnamefont{Pekker}}, \bibnamefont{and}
  \bibinfo{author}{\bibfnamefont{D.-W.} \bibnamefont{Wang}},
  \bibinfo{journal}{Phys. Rev. A} \textbf{\bibinfo{volume}{85}},
  \bibinfo{pages}{013603} (\bibinfo{year}{2012}).

\bibitem[{\citenamefont{Baranov et~al.}(2011)\citenamefont{Baranov, Micheli,
  Ronen, and Zoller}}]{baranov2011}
\bibinfo{author}{\bibfnamefont{M.~A.} \bibnamefont{Baranov}},
  \bibinfo{author}{\bibfnamefont{A.}~\bibnamefont{Micheli}},
  \bibinfo{author}{\bibfnamefont{S.}~\bibnamefont{Ronen}}, \bibnamefont{and}
  \bibinfo{author}{\bibfnamefont{P.}~\bibnamefont{Zoller}},
  \bibinfo{journal}{Phys. Rev. A} \textbf{\bibinfo{volume}{83}},
  \bibinfo{pages}{043602} (\bibinfo{year}{2011}).

\bibitem[{\citenamefont{Matveeva and Giorgini}()}]{matveeva2012}
\bibinfo{author}{\bibfnamefont{N.}~\bibnamefont{Matveeva}} \bibnamefont{and}
  \bibinfo{author}{\bibfnamefont{S.}~\bibnamefont{Giorgini}},
  \bibinfo{note}{arXiv:1206.3904}.

\bibitem[{\citenamefont{Liu et~al.}(1998)\citenamefont{Liu, Swierkowski, and
  Neilson}}]{Liu1998}
\bibinfo{author}{\bibfnamefont{L.}~\bibnamefont{Liu}},
  \bibinfo{author}{\bibfnamefont{L.}~\bibnamefont{Swierkowski}},
  \bibnamefont{and} \bibinfo{author}{\bibfnamefont{D.}~\bibnamefont{Neilson}},
  \bibinfo{journal}{Physica B: Condensed Matter}
  \textbf{\bibinfo{volume}{249-251}}, \bibinfo{pages}{594 }
  (\bibinfo{year}{1998}), ISSN \bibinfo{issn}{0921-4526}.

\bibitem[{\citenamefont{Klawunn et~al.}(2010)\citenamefont{Klawunn, Pikovski,
  and Santos}}]{klawunn2010}
\bibinfo{author}{\bibfnamefont{M.}~\bibnamefont{Klawunn}},
  \bibinfo{author}{\bibfnamefont{A.}~\bibnamefont{Pikovski}}, \bibnamefont{and}
  \bibinfo{author}{\bibfnamefont{L.}~\bibnamefont{Santos}},
  \bibinfo{journal}{Phys. Rev. A} \textbf{\bibinfo{volume}{82}},
  \bibinfo{pages}{044701} (\bibinfo{year}{2010}).

\bibitem[{\citenamefont{Volosniev et~al.}(2011)\citenamefont{Volosniev, Zinner,
  Fedorov, Jensen, and Wunsch}}]{volosniev2011}
\bibinfo{author}{\bibfnamefont{A.~G.} \bibnamefont{Volosniev}},
  \bibinfo{author}{\bibfnamefont{N.~T.} \bibnamefont{Zinner}},
  \bibinfo{author}{\bibfnamefont{D.~V.} \bibnamefont{Fedorov}},
  \bibinfo{author}{\bibfnamefont{A.~S.} \bibnamefont{Jensen}},
  \bibnamefont{and} \bibinfo{author}{\bibfnamefont{B.}~\bibnamefont{Wunsch}},
  \bibinfo{journal}{J. Phys. B: At. Mol. Opt. Phys.}
  \textbf{\bibinfo{volume}{44}}, \bibinfo{pages}{125301}
  (\bibinfo{year}{2011}).

\bibitem[{\citenamefont{Astrakharchik et~al.}(2007)\citenamefont{Astrakharchik,
  Boronat, Kurbakov, and Lozovik}}]{astrakharchik_07}
\bibinfo{author}{\bibfnamefont{G.~E.} \bibnamefont{Astrakharchik}},
  \bibinfo{author}{\bibfnamefont{J.}~\bibnamefont{Boronat}},
  \bibinfo{author}{\bibfnamefont{I.~L.} \bibnamefont{Kurbakov}},
  \bibnamefont{and} \bibinfo{author}{\bibfnamefont{Y.~E.}
  \bibnamefont{Lozovik}}, \bibinfo{journal}{Phys. Rev. Lett.}
  \textbf{\bibinfo{volume}{98}}, \bibinfo{pages}{060405}
  (\bibinfo{year}{2007}).

\bibitem[{\citenamefont{B\"uchler et~al.}(2007)\citenamefont{B\"uchler, Demler,
  Lukin, Micheli, Prokof'ev, Pupillo, and Zoller}}]{buchler2007}
\bibinfo{author}{\bibfnamefont{H.~P.} \bibnamefont{B\"uchler}},
  \bibinfo{author}{\bibfnamefont{E.}~\bibnamefont{Demler}},
  \bibinfo{author}{\bibfnamefont{M.}~\bibnamefont{Lukin}},
  \bibinfo{author}{\bibfnamefont{A.}~\bibnamefont{Micheli}},
  \bibinfo{author}{\bibfnamefont{N.}~\bibnamefont{Prokof'ev}},
  \bibinfo{author}{\bibfnamefont{G.}~\bibnamefont{Pupillo}}, \bibnamefont{and}
  \bibinfo{author}{\bibfnamefont{P.}~\bibnamefont{Zoller}},
  \bibinfo{journal}{Phys. Rev. Lett.} \textbf{\bibinfo{volume}{98}},
  \bibinfo{pages}{060404} (\bibinfo{year}{2007}).

\end{thebibliography}

\end{document}